\documentclass[11pt, oneside]{article}   	

\begin{document}
{\bf The Quantum Theory of Fields, Vol. I (Foundations) and Vol. II
(Modern Applications)}.
By Steven Weinberg. Cambridge University Press, Cambridge, U. K. 
xxvi + 609 pp. for Vol. I
(1995);
xxi + 489 pp. for Vol. II (1996).

\vskip 0.4in
\begin{quote}
``Men and Women are not content to comfort themselves
with tales of gods and giants, or to confine their thoughts to the daily
affairs of life; they also build telescopes and satellites and accelerators,
and sit at their desks for endless hours working out the meaning of
the data they gather. The effort
to understand the universe is one of the very few things
that lifts human life a little above the level of farce, and gives
it some of the grace of a tragedy.''
\vskip 0.2in
$\qquad$$\qquad\qquad$$\qquad\qquad$
{S. Weinberg, in {\em The First Three Minutes (1977).}}
\end{quote}

I have been invited to write a review of these two volumes
on the quantum theory of fields.
I consider it to be an honor to be so invited. 
My views are necessarily personal. Nevertheless, I hope the reader
shall find them of some use in gaining the perspective of what is
ahead of her if she is courageous enough to commit herself to this
Dirac--like monograph of over a thousand pages. 
Perhaps a little preparatory reading shall be of immense help
in this journey for her. Therefore, I shall
describe to you my own journey to these two volumes in the 
hope that you, if you  are a beginning student, shall find
these two volumes accessible when you first encounter them.

This long--awaited two--volume set by Professor Steven Weinberg
is a deeply personal and logical
perspective on the quantum theory of fields. I found many of these 
ideas, first, in  a 1974--75 handwritten notes
that Professor Weinberg distributed to his class at Harvard
and which Dr. Wayne Itano kindly sent to
me when later I was a graduate student at Texas A\&M. 
In addition, as I have noted
in my review of Professor Lewis H. Ryder's {\em Quantum Field Theory},
my first introduction to Professor Weinberg's ideas came via
Professor Ryder's book. I still think that 
Professor Ryder's book is an essential 
preparatory reading for a beginning student
before she embarks on her journey to these two volumes of
Professor Weinberg. 

Next, I read three early papers of Weinberg entitled
``Feynman Rules for Any Spin''
and many related papers. The first two papers
in the ``Feynman Rules for Any Spin'' series
 were
published in {\em Physical Review} in 1964, and the third, five
years later, 
in 1969, in the same journal. My own copies of these three
papers are now worn out from use, with ``p'' in ``Spin'' slashed
through by Professor Roger Smith --- a pun that, I am sure,
Dick Feynman would have enjoyed. To me these  still make a
good honest reading and I take pleasure that now and then I have
been able to find a small error in, or an appropriate extension
of, these works.

Professor Weinberg's ideas, as is always the case with the
original and ever--young minds, are not static. So it was not
too long ago that Professor Weinberg wrote: 

\begin{quote}
``A distinguished nuclear physicist asked me not too long ago
what I thought of a proposal to do an experiment on the scattering
of a nucleus (I forget which nucleus) with spin--2, which aimed
at finding out experimentally which relativistic wave equation that
nucleus satisfied. This is all wrong. If you go into the streets of
College Park and a passer--by asks you what is the Lorentz transformation
of a particle of spin--$2$, you do not have to ask him if he is referring
to a symmetric traceless tensor wave function,
or a wave equation belonging to some other representation of the
homogeneous Lorentz group that contains spin--2. All you have to do is
$\ldots$ tell him that the states transform according to $j=2$ matrix 
representation of the of the rotation group (I promise you that if you do that
the pedestrian will not
ask you any more questions). The kinematic classification of particles
according to their Lorentz transformation properties is entirely
(for finite mass) determined by their familiar representation of the
rotation group. It has nothing whatever to do with the choice of one
relativistic wave equation rather than another.''

\vskip 0.2in
$\qquad\qquad\qquad$
 S. Weinberg, {\em Nucl. Phys. B (Proc. Suppl.)} {\bf 6}, 67 (1989).
\end{quote}
The background for this argument is contained
in the beginning chapters of the ``Foundations.''
It is not clear to me if Professor Weinberg still holds this
view without any qualifications. For one thing, recently it has become
possible to construct a field theory in which a $(1,0)\oplus(0,1)$
boson and its associated anti--boson carry {\em opposite} relative intrinsic
parity. In the standard description, in terms of a $(1/2,1/2)$ 
representation,   spin-$1$ bosons and anti--bosons 
have the {\em same} relative intrinsic parity.
Of course, 
whether the boson--antiboson pair carries the same, or opposite, 
relative intrinsic parities
has experimental consequences.  Generalizations
of this argument exist for spin--$2$ and higher. 
Moreover, interactions that are ``simplest'' in one representation may look
complicated in another, as Professor Weinberg himself notes in 
Sec. 5.7.

In 1962 Professor Wigner, in  a collaborative work with
Professors V. Bargmann and A. S. Wightman, suggested the 
possibility that unusual P, and CP, properties of particles
and antiparticles may exist within the kinematic considerations
that began with his 1939 work. However, no explicit constructs
for these unusual Wigner-- (or ``BWW--'') type
field theories  existed prior to
1993. In this context Professor Weinberg (see p. 104) notes
that ``No examples are known of particles that furnish
unconventional representations of inversions, so these 
possibilities will not be pursued further here.'' In my opinion,
it is not yet clear what  the general phenomenological
and theoretical
implications of these newly discovered constructs are. The  
experimentalists, and entering graduate students,
should be made aware of these unusual possibilities --- or, so atleast
is my opinion.

The structure of quantum field theory is exceedingly rich.
This becomes abundantly clear as one reflects on the 
presentation of Professor Weinberg. For instance,
at least a generation of physicists simply wrote down the
spin--$1/2$ rest spinors {\em a la} Bjorken and Drell.
In Professor Weinberg's monograph the same result is obtained
after  reflections and calculations spanning about five pages
(see Sec. 5.5). I refer the reader to hep-th/9702027, written
by Professor Weinberg under the title
``What is Quantum Field Theory, and What Did We Think It Is?''
for a very readable essay on the logical structure of 
the theory of quantum field theory. In the cited essay
Professor Weinberg writes, ``This has been an outline of the way
I've been teaching quantum field theory these many years. Recently
I've put this all together into a book, now being sold for a 
negligible price. The bottom line is that quantum mechanics
plus Lorentz invariance plus cluster decomposition implies quantum field 
theory. But there are caveats that have to be attached to this  $\ldots$. ''
It is not appropriate for me to comment on other chapters of these
monographs --- even though almost all of them carry
many original arguments and derivations ---  for the simple reason
that many others (much more knowledgeable than I on these chapters) 
have already done
so elsewhere in various reviews already published. Some may find it 
a little strange
that a monograph of this nature contains no mention of theorists 
such as R. Arnowitt, J. J. Sakurai, or E. C. G. Sudarshan, to mention just
a few. The origin of such omissions, in my opinion, may lie in the fact
that this two--volume monograph is a deeply personal perspective
of its author.

In Professor Weinberg's {\em Gravitation and Cosmology} some of us
learned
how the empirically observed equivalence of the inertial and
gravitational masses yields Einstein's general theory of relativity.
In that book the starring role was taken away from the
geometric elements and was rightfully transferred to the 
experimental facts.
Now we learn how quantum mechanics, space--time symmetries, and 
the requirement that spatially separated experiments yield
uncorrelated results, intermingle in such a manner as to result
in a quantum theory of fields.
We may be approaching something close
to the Landau and Lifshitz classics inasfar as two of the most
important aspects of modern physics are now covered by one single 
author of relentless effort and uncompromising
honesty. 

In the preface to the second volume
Professor Weinberg notes ``Perhaps supersymmetry and supergravity will be 
the subjects of a Volume III.''
Before the Weinberg trilogy  is completed, I hope we all have had
the opportunity to study, examine, and enjoy Professor Weinberg's
{\em The Quantum Theory of Fields} in great detail.
This work compares in its depth with P. A. M. Dirac's
{\em The Principles of Quantum Mechanics} and without doubt
goes far beyond the ambition of providing a quick calculational
recipe book. 
It is my opinion that
{\em The Quantum Theory of Fields} is a 
serious scholarly attempt to lift  
``life a little above the level of farce.'' It would be 
a tragedy,
if the expediency of finishing Ph.D.s and making calculations made
{\em The Quantum Theory of Fields}  a bookshelf decoration.

\rightline{D. V. Ahluwalia$\,\,\quad\qquad\qquad\qquad$}
\rightline{\sl Los Alamos National Laboratory}
\rightline{\sl Los Alamos, NM 87545, USA.$\,\,\quad$}

\vspace{21pt}

\noindent
\textsc{A note added on 09 July 2023}
\vspace{7pt}

In the second paragraph of this review I quoted Weinberg asserting that, ``No examples are known of particles that furnish
unconventional representations of inversions, so these 
possibilities will not be pursued further here.''  I can now safely assert that hidden in the ``unconventional representations'' lies a field theory that provides a natural dark matter candidate. The reference is: D. V. Ahluwalia, J. M. Hoff da Silva, C.-Y. Lee,
Nuclear Physics B
Volume 987, February 2023, 116092

\end{document}